\documentclass[11pt,a4paper]{article}
\usepackage[hyperref]{acl2020}
\usepackage{times}
\usepackage{latexsym}

\usepackage{graphicx}
\usepackage{amsmath}

\usepackage{microtype}
\usepackage{tabularx}

\aclfinalcopy 

\title{Consistent, Central and Comprehensive Participation on Social Media}

\author{Julian Dehne \\
and Valentin Gold
  \\}

\graphicspath{ {./images/} }

\date{}

\begin{document}
\maketitle
\begin{abstract}
Participation research in online community has concentrated on popularity and social interactions. In this paper the attention is shifted to the conversational trees as the focus of analysis in order to achieve a measure of deliberative participation. 
Three methods to measure consistent, central and comprehensive participation (CCCP) in online conversations are proposed. 

\end{abstract}

\section{Introduction}

Measuring participation in online communities has seen a lot of research and is based on the mental model of a shared space where community members engage. \cite{preece_sociability_2001} differentiates between communities based on their interactivity, the usage, the reciprocity and many other determinants. 

\cite{malinen_understanding_2015} reviews the research about participation in digital media and social networks. They come to the conclusion that it has been "operationalized mainly in terms of its quantity. The most commonly employed quantitative measures include duration of membership, time spent online, number of visits, number of hits/views of content, number of contributions, and density of social interaction with others". From this list of research activities a measure that summarizes consistency, centrality or comprehension is missing. This is the gap that is being addressed here. 

Intuitively, measuring comprehensive and consistent participation asks the question whether learning processes have a chance to take place. While a fleeting look might give a first impression, more involvement makes it more likely that arguments or other perspectives take hold long enough to change the behaviour in the offline world or at least in during the online conversation. 

When looking at prominent users like politicians the deeper involvement in conversations their consistent and comprehensive participation also sheds a light whether or not social media are used for marketing only. 

In this paper the unit of analysis is the conversational structure defined by the reply trees of online conversations. This smaller focus has a number of advantages: 

\begin{itemize}
    \item it is agnostic to the platform (generality)
    \item it can be transferred to offline conversations (medium-agnostic)
    \item it does not require complex text analysis \\ (efficiency)
    \item it does not abstract away from single users (multi-purpose)
\end{itemize}

The explanatory power of the conversation structure is that it shows the intensity of user engagement in terms of their comprehension of the previous arguments, the position of their contribution and the generality of these features. 

If a general measure of central and consistent participation could be established, platforms could then be compared regarding the presence of leaders that shape a conversation in a comprehensive (overall) way.

As a first step to identify these participation modes, the conversation structures are analyzed in order to identify reply trees that qualify as discussions (rather than broadcasts) and to also identify authors that may play a bigger role than a fleeting participant.


\section{Related Work}

\citep{bakshy_everyones_2011} and \citep{magnani_conversation_2012} both introduce user-centric measures but focus on popularity or influence of the user's tweets.

\cite{joglekar2020analysing} define responsiveness, reciprocity, branching factor and centrality as measures of describing the author roles in a reply tree. Using a Reddit forum with a high degree of repeated user interaction, the in-between-centrality can be used to measure the centrality of authors in these cross-conversational interactions. Using the cross-conversational user interactions as the unit of analysis for the centrality computation does not work in the context defined in this paper as both Reddit and Twitter data is analysed with the Twitter data being much sparser in direct user interactions.

\cite{aragon2017thread} discuss generative models of online discussion threads. The prediction based algorithm that is introduced later builds on the idea that online discussion can be viewed as bipartite graphs with on the one side the reply trees and the on the other side the author interaction graph. Similar to the authors it is assumed that the interaction structure and the reply tree both have explanatory value even if the text content is disregarded completely. Moreover, the author interaction graph can be interpreted as the visible part of a hidden Markov chain: whether it be following structures, common interests or similar access to the same links in Twitter or Reddit, for all these variables and more, cultural-social aspects may influence who answers to whom and why. But for these predictive models that are focused here, the only importance is the fact that these users enter into a conversation and even deeper dialogues. 


\section{Theoretical Framework}

Conversations are defined as a reply-tree with the original post or tweet being the root. ~\citep{cogan_reconstruction_2012} use a similar definition to analyze different conversation structures in Twitter. Since their paper Twitter is now providing an identifier for conversations based on their own algorithms. This serves as a starting point. However, conversation may contain up to thousands of posts. As the target size of a conversation should be comparable to a human conversation, trees with more than 100 nodes are excluded. At a later stage segmenting larger trees is planned, too.

~\citep[360]{magnani_conversation_2012} define conversations as a set of polyadic interactions with the following properties

\begin{itemize}
    \item a conversation is a set of interactions
    \item each interaction consists of two posts
    \item the interactions form a tree
    \item the timestamps give a strict order for all nodes in the conversation tree 
\end{itemize}

More relevant than the definition of the conversation tree is the concept of \textsl{scope}. Nodes have the property that they belong to an author who wrote the post. When writing this author had the scope of all the posts that were written in the same conversation before the one under inspections. 

The concept of consistent, central and comprehensive participation (CCCP) that is developed in this paper uses the \textsl{scope} as a necessary basis of the thought experiment but not a sufficient one. Although theoretically the author could have answered all of the posts in the conversation beforehand he would have to have read them which is not certain but more and more unlikely the bigger the conversation is and the further apart previous posts are. Comprehensive participation means that the participants has a good and thorough understanding of what was written before joining in. Consistency means that the participants take part in as many as possible branches of the conversation.

\begin{figure}[htp]

\centering
\includegraphics[width=.3\linewidth]{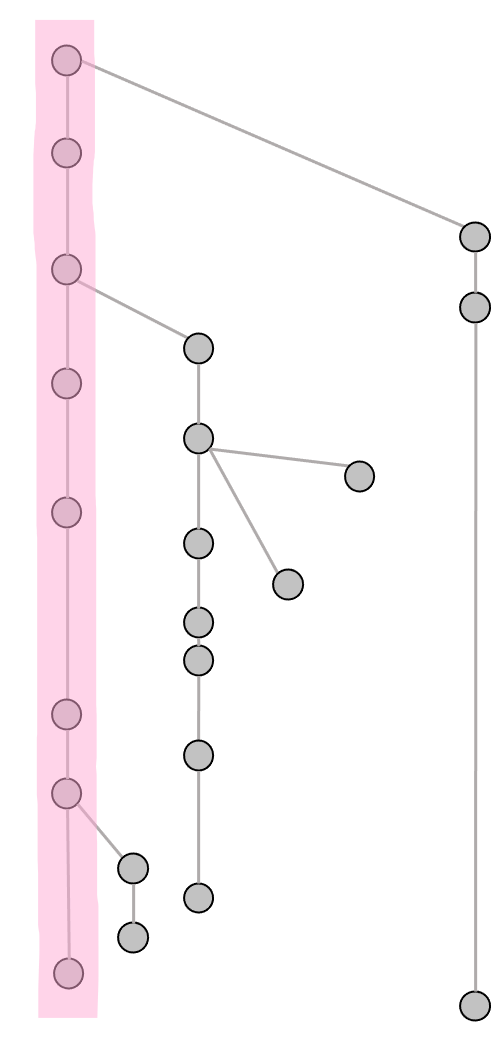}\hfill
\includegraphics[width=.3\linewidth]{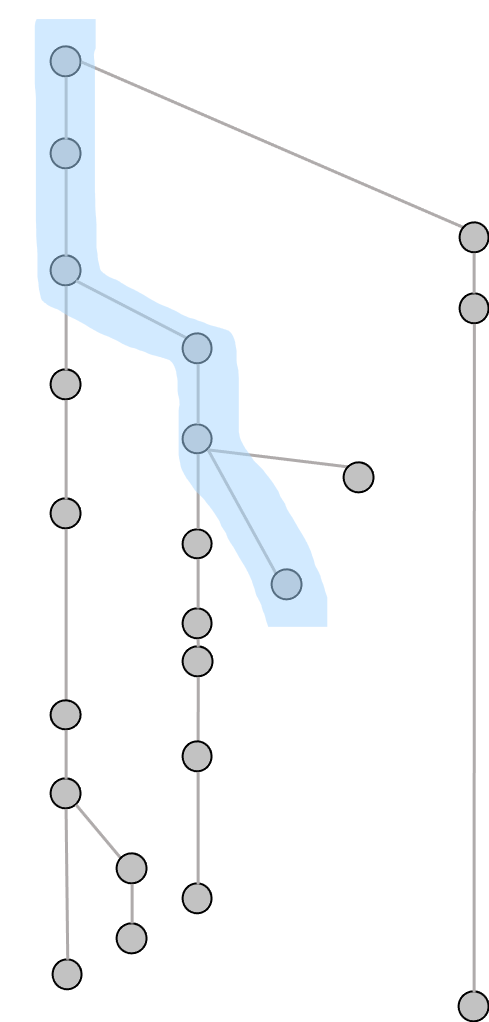}\hfill
\includegraphics[width=.3\linewidth]{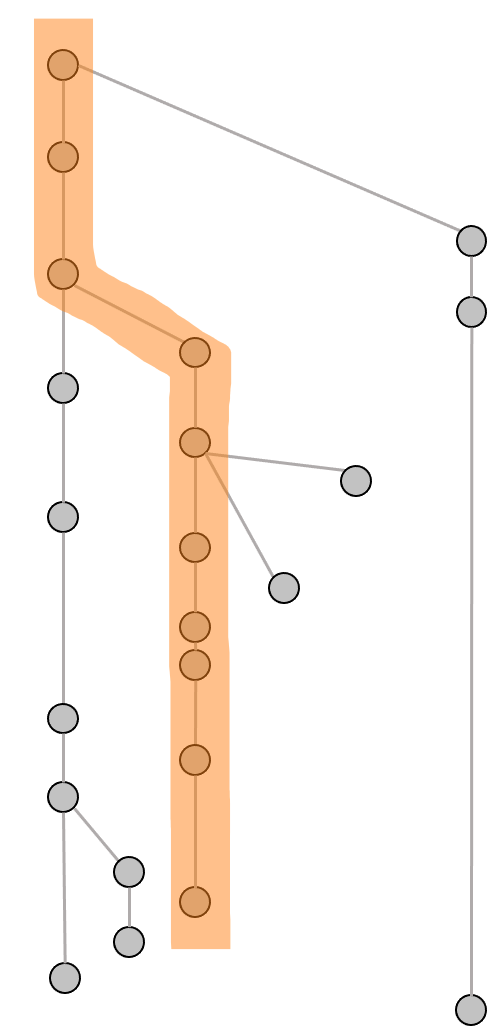}

\caption{Conversation flows within the reply tree: if the trees are viewed as as list of branching linear conversations it is illustrated in pink the first flow, in blue a second and in orange a third. More flows exist in this tree.}
\label{fig:figure3}

\end{figure}

In a live discussion guessing who might fulfill the role of an active participant would start by noticing the number of verbal interactions the different participants have. The analogy breaks with the simultaneity of different discussion branches in the online environment which cannot be mimicked in a live discussion. Although it could be possible to create breakout-rooms as equivalents to the branching of online conversations, no participant could meaningfully be active in all breakout rooms at the same time. In an online settings this is possible although not very likely either.   

\begin{figure}[h!]
    \centering
    \includegraphics[width=\linewidth]{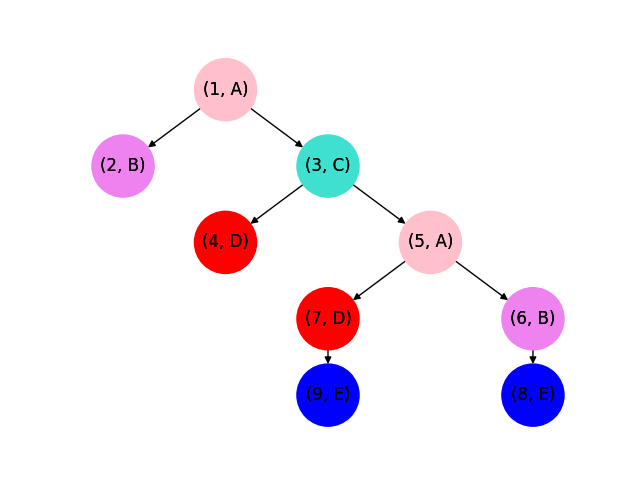}
    \caption{Example Conversation Tree. The integers are the node identifier and the letters and colors represent the authors writing the posts}
    \label{fig:example_tree}
\end{figure}

The example tree in figure \ref{fig:example_tree} will be used to illustrate the different approaches to modelling CCCP and author centrality. It includes a minimal set of features that are available for any conversation. It should be noted that platform specific features like follower counts, mentions or popularity measures like up-votes are excluded. The advantage of using this reduced set of features is that it can be easily generalized to even include printed interviews or recorded live discussions.

The participant with the highest CCCP score can be intuitively considered a moderator or discussion leader. In order to make sense of the CCCP concept using the moderator metaphor is helpful.

Given the example tree there is no clear candidate for a moderator as most authors have two contributions and even the author with one contribution has uttered a comment at a very central part of the tree. With the definition of scope based on ~\citep[360]{magnani_conversation_2012} the blue author (E) has the highest scope. However, the nodes 1 and 3 are far away from the posts and assuming a decay function on the path distance of the reply tree, the pink author may have much higher probabilities of having seen these nodes. The intuition dictates that a combination of these aspects should be used as a metric for CCCP. 

In contrast, it seems very clear that the beige author (A) is the most central in the discussion. He is a part of all paths in the tree that begin at the root node and end in a leaf. These paths play a specific role as they can be viewed as the sub-conversations that are equivalent to the analog version of a conversation. According to ~\citep[359]{magnani_conversation_2012} these can be modeled as a \textsl{"quasi-chain structure with a rigid chronological sequence of interactions where almost every message refers to the previous one, as it usually happens with off-line conversations"}. These paths will be referred to as conversation flows. Moreover, for the active CCCP role one would assume that the user would speak distributively in the conversation and mostly in a central position within the conversation flows. 

In the following these intuitions will be expressed a mathematical formulas and algorithms for computing corresponding metrics for larger data-sets will be presented. 

\section{CCCP Metrics}

The comprehensive participation means comprehensive in a cognitive way as well as a quantitative attribute. Having seen and processed more of the former posts, the participation will be more active and informed.

Given a general tree like the one in the example (figure \ref{fig:example_tree}) there is little information in order to estimate the likelihood of an author having seen a post having previously written a number of posts in the conversation. 
For a given conversation flow these pair-wise likelihoods of a later post's author having seen the earlier post is defined as: 

\begin{multline}
    seen(V_{j}, A_{k}) = \bigcup_{V_{i}\in\varphi}(P(SEEN\mid(V_{j},V_{i}))) \\
    \vee isAuthor(A_{k},V_{i}) = \top \vee V_{j}\in\varphi
    \label{eq:seen}
\end{multline}

Equation \ref{eq:seen} reads as: the probability that an author has seen the post j is the union of the SEEN-probabilities of pairs of nodes in the reply tree $\varphi$. 

We found three approaches for modelling the SEEN property.

\begin{enumerate}
    \item an author has seen a post depending on its distance from the root post combined with the distance of the next answer in the reply tree of the same author
    \item an author has seen a post for sure if and only if he has answered it. These positives are used to train a neural network that trains weights for the two distances used in the previous approach
    \item an author has seen a post if he can be predicted to answer it given the previous conversation
\end{enumerate} 

These different approaches each resulted in their own algorithm that try to give a number to this idea. The first is used as the baseline as it can be modeled and computed directly without any complicated machine learning. The second and the third approach are implemented as classification and prediction tasks.
 
\subsection{CCCP Metric Baseline} 

The baseline uses to two assumptions mentioned previously: an author has seen a post with a probability depending on its distance from the root post combined with the distance of the next answer in the reply tree of the same author. 

\begin{multline}
    \zeta := P(SEEN\mid(V_{j},V_{i}))) = \\ 1/(\mid\iota\mid)\sum(1/2)^{(\mid path(V_{j},V_{i}) \mid-1)} 
    \label{eq:pathdistance}
\end{multline}

The equation \ref{eq:pathdistance} reads as: the probability zeta of having seen node j is the average sum of the decay function of the path length between all the nodes i written by the given author (see equation \ref{eq:seen}) and the node j. For example if the author has only written one subsequent answer to a post and this answer has a path distance of two replies to the post j than the probability of having seen j for the author would be $1/2^{(2-1)}=0.5$. If the path distance was 1 for a direct reply the exponent would be 0 and the probability 1. This measure is computed as an average for all existing path between node j and nodes i of the given author. Analogously, the root distance can be defined as

\begin{multline}
    \vartheta := P(SEEN\mid(V_{j},V_{i}))) = \\ 1/(\mid\iota\mid)\sum(1/4)^{(\mid path(root,V_{j}) \mid-1)} 
    \label{eq:rootdistance}
\end{multline}

\begin{equation}
    \Rightarrow P(SEEN) = \zeta \cup \vartheta
\end{equation}
 
 The advantage of defining the author metric with this rule based approach is that it can be computed quickly for each conversation in parallel without previous training and fitting. However, this approach can only function as a baseline as it models the assumptions (that the root distance and the reply distance matter) without providing empirical prove for the latter or for the exact numbers used as for the bases of the decay functions. 
 
\subsection{CCCP based on Direct Responses (RB)}

The naive approach of modelling the assumptions with assumed weights and decay functions can be improved by using safer assumptions to train weights for the weaker ones. One such strong assumption is that an author that has replied to a post has actually read it. This way we can transform the CCCP metric to a machine learning problem. Positive training examples are tweets that have been answered by the given author. Negative examples are the rest. As features the reply distances are used as well as the root distances and the time deltas between the posts. 

Not only avoids this approach some of the ambiguity of the baseline model but it also enlightens the latter as the embedding can be interpreted as the weights used for the decay functions. If the neuron for the feature reply-distance-1 is .9 and the neuron for the feature reply-distance-2 is loaded with 0.45 after training, this would suggest a basis of (1/2) for the decay function.

Using the example graph from figure \ref{fig:example_tree} the pairs that would produce a y = 1 for training are:

\begin{table}[h!]
    \centering
    \begin{tabular}{c|c|c|c}
     author & $V_{i}$ & $V_{j}$ & y \\ \hline 
       A    & 5  & 3  & 1 \\
       B    & 6  & 1  & 1 \\
       B    & 6  & 5  & 1 \\
       C    & 3  & 1  & 1 \\
       D    & 4  & 3  & 1 \\
       D    & 7  & 5  & 1 \\
       D    & 7  & 3  & 1 \\
       E    & 9  & 7  & 1 \\
       E    & 8  & 6  & 1 \\
    \end{tabular}
    \caption{Preparing the training labels for the CCCP algorithm based on Direct Responses}
    \label{tab:my_label}
\end{table}

Training with a structure like this example conversation would give the node D the highest score (instead of E which was the intuition). 

\subsection{CCCP based on Author Predictions (PB)}

The third idea for calculating whether or not an author has seen a node is by taking a step back and artificially forgetting which author has written the last post for each post in turn. This is a typical approach in machine learning: by inferring the current information from the context something can be learned about the context. In this case we assume that the information whether an author takes part in the previous conversation is encoded in the context and answering in the current situation would suggest that an author has read more of the conversation. If an author does not answer he might have read less. Whereas the previous approach was backward-looking this inverts the machine learning question to a prediction task: the sum of the probabilities of predicting a given author for each situation in the conversation is assumed to correlate to the overall vision the author had at the time of writing.

The semantics of this approach include many aspects from \cite{aragon2017thread}: the probability of an author writing is dependant on the times she has written previously and the general likelihood that a new author would join the conversation. The execution of this calculation differs as it includes other aspects like the social network with the previous authors (following/not following), the structure of the reply tree. Using the interaction tree between the authors, too, could be considered in future implementations.

In terms of formulas this approach is very close to the one taken by \cite{aragon2017thread} and for these reasons the formulas don't need to restated here. The main difference is that the main goal is not to predict author participation but invert the question and ask whether the user has been active enough in order to be predicted to write next. Because of the sparser user interactions the interaction graph had to be removed from the equation resulting in a rather under-defined structure. As can be seen later, significant differences between the platforms can still be seen.

Predicting the new author on the structure alone seems very unlikely. But this approach benefits from a general perspective on the path distances between the nodes. Indeed the only assumption it makes is the fact that the reply structure implies which author might write next. 

Given that the sample cuts off longer conversations a low precision is to be expected. Longer samples inform the algorithm more, because they would have more data points per conversation. Consequently, ignoring longer conversations for computational reasons (computation costs rise by $ 2^{len(c)}  $), reduces the validity of the results. Despite running these algorithms on a cluster, a max length of 100 posts per conversation flow had to be imposed. This seems acceptable as it can be assumed that very few users read 100 posts or more.

\section{Comparing CCCP on Reddit and Twitter }

In order to evaluate the newly developed metrics no existing baselines could be used. For this reason a double quasi-experiment was chosen comparing the results for Reddit and Twitter on the one hand and between the different approaches on the other hand. The metric would be evaluated as stable if it worked the same for Reddit and Twitter and as precise if the different approaches would lead to results that are correlated.

\subsection{Sampling}

The following table shows the samples drawn from Twitter and Reddit:

\begin{table}[h!]
    \centering
    \begin{tabularx}{\linewidth}{l|l|X}
        Variable & Reddit & Twitter \\\hline \hline
        n conversations large & 1229 & 4240 \\
        n posts large & 60881 & 372567 \\
        \hline
        n conv. large resampled & 546 & 546 \\
        n posts large resampled & 514793 & 535271 \\
        \hline
        n conversations small & 1195 & 34 \\
        n posts small & 76481 & 3085
    \end{tabularx}
    \caption{Drawn Sample from Social Media}
    \label{tab:sample_size}
\end{table}

The conversations were downloaded using the APIs of the platforms respectively. Only conversations that could be fully downloaded without deleted posts or missing parents in the tree structure were used in sample. In order to find more trees that contain a discussion and not a broadcast or advertisement, topics like climate change, vaccination or immigration were used. In theory, the topics should not change the quality of the abstract conversation metrics too much based on the high number of samples drawn. In order to judge the influence of the sample size both a large and a small sample were drawn independently with different search queries with one sample leaning towards more Reddit data and the other leaning towards more Twitter data. In the case of the larger data-set the data was re-sampled in order to have the same number of conversations for both platforms.

\subsection{Implementation of the Algorithms}

The baseline formula does not contain any computation complexity so it will not be explained in detail beyond the formulas given in equations \ref{eq:seen},\ref{eq:pathdistance} and \ref{eq:rootdistance}.

For the backward looking algorithm a neural network was trained with one hidden layer using stochastic gradient descent as optimizer and binary cross-entropy as a loss function. The trained model was applied to all data points and the average sum of the paired vision probabilities were computed grouped by the conversation and the platform. 

For the forward looking author prediction algorithm a neural network with six dense layers and a softmax layer was used. As this was a multi-label classification categorical cross-entropy was used. The options for authors to be predicted were generated using one-hot encoding. In order to compute the classification whether a new author can be predicted or one of the existing ones an extra column was generated that was 1 if the author was writing for the first time in the conversation. The actual author column was than set to 0 in order to have a categorical classification. The neural network would than either output the new author category or the specific author that was most likely to answer given the context.

\subsection{Empirical Results}

The following table shows the CCCP for Reddit and Twitter grouped by platform, conversation and author using the mean as the aggregation function. 

\begin{table}[h!]
    \centering
    \begin{tabular}{c|c|c|c}
        Algorithm & Data-set & Reddit & Twitter\\ \hline 
        Baseline & big & 0.518163 & 0.466651\\ 
        Baseline & small & 0.436952 & 0.135928   \\ 
        RB & small & 0.229424 & 0.096224 \\
        RB & big & 0.198840 & 0.169384 \\
        PB & small & 0.607303* & 0.400219* \\
        PB & big & 0.431931* & 0.170356* \\
        Centrality & small & 0.41897** & 0.26566** 
    \end{tabular}
    \caption{Comparing different CCCP metrics for Reddit and Twitter.\newline *the prediction based numbers were divided by the repetition probabilities in order to make them comparable \newline **the author centrality is computed as an adapted in-between-centrality based on the conversation flows}
    \label{tab:results}
\end{table}

\bigskip
It is noteworthy that for all the different measures there seem to be stable results when comparing the platforms. However, the precision for the response based algorithm was 98~\% whereas the precision for the prediction algorithm only amounted to 47~\%. It should also be noted that 80\% (or higher) of the contexts were predicted to have a new author writing the next posts. The precision is much lower if the predicted authors are aligned without introducing the "new author" category or without normalizing with the repetition probabilities per conversation.

Even more surprising is the fact that the author centrality also aligns with the CCCP. This could suggest a empirical link between having a good overview over a discussion and being in a central position. However, a cross-method comparison that correlates the two measures on the basis of conversations edit distances would be useful to shed light on this relationship. 

When inspecting the embedding generated by the RB algorithm it was surprising that the weights for the distances did not follow a decay function at all. This is due to the oversight that someone is not very likely to answer herself. This leads to a set of interesting series of decays based on the tree structures and the length of the longest path. Future work should investigate segmenting these series of decays into shards that can be normalized by the repetition probabilities. 

Although aggregated for the platform the measures seem to indicate the same trends, individual conversations differ in their scores indicating that the measures do indeed measure different concepts.

\begin{figure}[h!]
    \centering
    \includegraphics[width=\linewidth]{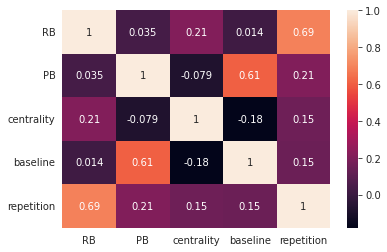}
    \caption{Correlation with the Pearson method does not show significant relationships between the measures}
    \label{fig:correlations}
\end{figure}


In all the algorithms the closeness to the root played a more important role than the reply distance supporting the existing theory of preferential attachment. 

A major issue with this investigation is the lack of a gold standard. The different measures focus on different aspects of CCCP, the baseline and RB-algorithm on comprehension, the PB on consistency and centrality is calculated separately in any case. For this reason the correlation table should be viewed a proof of the internal consistency of the measures whilst holding up the categorical independence. 

As a way forward these measures could be used together to create a meta-measure that takes into account all aspects of CCCP. Another suggestion would be to use platform dependant information like author interaction, author networks or more to cross-validate the measures. 

For the less-technical oriented computational social science community the repetition probabilities combined with the centrality might be a good starting point as both are tried and tested and do not require training a model first.  

\section{Discussion and Summary}

Using a generic conversation model it was shown that there are platform dependent differences between author presences in a discussion. It is likely due to the stronger group adherence in Reddit that four different experimental measurements show the same pattern of higher involvement spread out through the conversations. 

The measures reflect different interpretation of CCCP (backward-looking, predictive, centrality-based etc.) and could be shown to be independent. Given that they give a stable prediction for the platforms they could be used in combination as an integrated CCCP index. As an alternative the simple to compute baseline could be used to approximate the CCCP and thus provide a means for the less technology oriented community to investigate author involvement in social media.

The assumption that only the distance to the original post or the length of the reply trees are relevant for the CCCP is necessary but problematic. In both platforms there are ways to react cross-branch using @mentions or in the case of Reddit by copying links. However, there is no conclusive way to compare these different structures. 

It is planned in the future to use platform specific knowledge like follower networks to add features to the PB algorithm in order to improve the precision and to the RB algorithm in order to improve the sensibility. Although this looses the cross-platform comparability it could inform some of the current questions surrounding the differences compared to the baseline.

Although the main goal is to find user moderators in the social networks, the above shown metrics can be used in a number of other applications:

\begin{itemize}
    \item Investigating deliberative quality in social media based on author involvement
    \item Compare different language areas based on their discussion leaders
    \item Investigate the involvement of politicians after their initial post on a social media platform
\end{itemize}


\bibliography{delabbib,author_vision_1}
\bibliographystyle{acl_natbib}

\end{document}